\documentstyle{article}

\begin{document}

\title{Schwinger terms in 1+1 dimensions}
\author{Tom\'a\v s S\'ykora \\
Department of Nuclear Centre \\
Faculty of Mathematics and Physics, Charles University \\
V Hole\v sovi\v ck\'ach 2, 182 00 Prague, Czech Republic}
\date{July 13, 1997}
\maketitle

\begin{abstract}

Two different approaches - K\"allen's and Brandt's methods - for
calculation of the Schwinger terms in the 1+1 dimensional Abelian and
non-Abelian free current algebras are discussed. These methods are
applied to calculation of the single and double commutators.
The validity of the Jacobi identities is examined in 1+1 and 3+1
dimensions and in this way is given natural restriction on the
regularization. It is shown that the Jacobi identity cannot be broken
in 1+1 dimensions even using the regularization which fails in the 3+1
dimensional case.
A connection between the Schwinger term and anomaly is shown in the
simplest case of the Schwinger model.
\end{abstract}



\section{Introduction}


\label{1} \renewcommand{\theequation}{1.\arabic{equation}} %
\setcounter{equation}{0}

Schwinger terms \cite{Sch1,Sch2,Goto}, i.e. terms proportional to the
derivatives of the $\delta$-function in commutators of currents or
gauge generators, play an important role in the investigations of various
field theory models \cite{Sto}-\cite{Be1}.

They are closely related to quantum anomalies and therefore to the
completeness and the self-consistency of a given quantum field theory
(QFT). This is the main reason they should be considered.

In particular, it is well known that the seagull terms
which are introduced to covariantize the time ordered product of current
operators do not cancel the Schwinger terms in the equal time commutator
(ETC) and Feynman's conjecture fails \cite{Gross}.

In this paper I will present a r\'esum\'e of the results which
can be obtained using two different approaches for the calculations
of the Schwinger terms in the 1+1 dimensional Abelian and non-Abelian
free current algebras.

Roughly speaking, if the algebra of the currents or the gauge
generators \cite{Sto}-\cite{Be1} of the quantum theory is not closed we
get {\it an indication} this theory could be the anomalous and then
the whole process of its quantization is highly non-trivial
\cite{Jack}-\cite{Il}.

There are several methods for the calculation of the ETC.
Here I use K\"allen's \cite {Kal} and Brandt's \cite{Bra}
methods. From my point of view these two methods are more elegant and
simple, especially in 1+1 dimensions, than e.g. the BJL-method
\cite{Bj,Lo}. I think the main difference lies in the {\it logic} of
the problem.

In K\"allen's or Brandt's methods we start with the canonical
commutation relations, then we calculate the ETC and finally the anomaly
(see Section 4).

In the case of the commutator of the two currents using BJL-technique
we have to calculate the vacuum expectation value (VEV) of their
T-product, i.e. all relevant Feynman graphs. Only after this we are
able to find the anomaly and, subsequently, the ETC. This
is obviously a rather indirect procedure.

Of course I am aware of the problems of Brandt's and K\"allen's method
in 3+1 dimensions.
Brandt's method fails to give reliable results once we get to the
spatial limit (e.g. \cite{Bra}).
Problem of K\"allen's method lies in the possibility of
changing the order of the integration (see (\ref{valid})).
Nevertheless I believe that it is possible to improve these methods.

Another open question, which is closely related to equal-time algebras, is
a violation of the Jacobi identity. Again, it is possible to use
BJL \cite{Levy}-\cite{Wud}, K\"allen's or Brandt's (point-splitting)
\cite{Ban} methods.

The key point of the problem consists in correct definitions of
current operators and their multiplication.

In general, it would be desirable that the Jacobi identity is
fulfilled, since the multiplication of {\it well-defined operators
on the Hilbert space} is necessarily associative.
Regularization procedures defining the current commutators and
respecting the associativity (i.e. the Jacobi identity) do exist
(see Sect. VII); therefore they have to be considered as prefered
(correct) ones.

For compact reviews which contain similar topics see \cite{Adam}-\cite{Paw3}.

This paper is organized as follows. In Sect. II resp. III the
definitions of the ETC according to K\"allen
\cite{Kal} resp. Brandt \cite {Bra} are given, the single commutators
are calculated using both methods and the results are used in Sect. IV
for the calculation of the anomaly.

In Sect. V resp. VI the double commutators are computed using the
K\"allen's and Brandt's techniques and the results are used in
Sect. VII for the discussion of the failure of the Jacobi identity in
1+1 and 3+1 dimensions.


\section{K\"allen's approach}


\label{2} \renewcommand{\theequation}{2.\arabic{equation}}
\setcounter{equation}{0}
The technique employed in this section was introduced by K\"allen \cite{Kal}.
\\Let us consider the function
\begin{equation}
F^{ab}_{\mu\nu}(x-y)=
\langle 0| J_{\mu}^{a}(x)J_{\nu}^{b}(y) |0\rangle,\label{K1}
\end{equation}
where $\langle 0|\dots|0\rangle$ means the VEV and
\begin{equation}
J^{a}_{\mu}(x)\equiv\bar{\psi}(x)\gamma_{\mu}\tau^{a}\psi(x),
\end{equation}
where $\gamma_{\mu}$ and $\tau^{a}$ are the Dirac and Pauli matrices
respectively.

Inserting the completeness relation between the two currents in (\ref{K1}) we
obtain \cite{Itz}, 
\begin{eqnarray}
F_{\mu \nu }^{ab}(x-y) &=&\sum_n\langle 0|J_{\mu}^{a}(x)|n\rangle \langle n|J_{\nu}^{b}
(y)|0\rangle   \nonumber \\
&=&\sum_n\langle 0|J_{\mu}^{a} (0)|n\rangle \langle n|J_{\nu}^{b} (0)|0\rangle
\: e^{-ip^{(n)}(x-y)}  \nonumber \\
&=&\int {d^2p\:e^{-ip(x-y)}G_{\mu \nu }^{ab}(p)\theta(p^{0})},
\end{eqnarray}
where the summation runs over the states with positive energy and
\begin{equation}
G_{\mu \nu }^{ab}(p)= \sum_{\sum_i{p_i^{(n)}}=p}\delta(p^{(n)}-p)\langle 0|J_\mu
^a(0)|n\rangle \langle n|J_\nu ^b(0)|0\rangle .  \label{G}
\end{equation}
The many-particle state $|n\rangle $ has total momentum $p$. $G_{\mu
\nu }^{ab}(p)$ vanishes for $p^2<0$ or $p^0<0$.

We can find $G^{ab}_{\mu\nu}(p)$ using several methods. The simplest one
is to notice that $G^{ab}_{\mu\nu}(p)$ must be a tensor of rank two with
respect to Lorentz transformations and therefore its most general form is
\begin{eqnarray}
G^{ab}_{\mu\nu}(p) & = & A^{ab}(p^{2})g_{\mu\nu} +
B^{ab}(p^{2})p_{\mu}p_{\nu}  \nonumber \\
& = & \bigl(p^{2}g_{\mu\nu}-p_{\mu}p_{\nu}\bigr)G^{ab}_{1}(p^{2}) +
g_{\mu\nu}G^{ab}_{2}(p^{2}).  \label{GG}
\end{eqnarray}
If the vector current $J_{\mu}$ is conserved, i.e.
$\partial_{\mu}J^{\mu}=0$, then $G^{ab}_{2}(p^{2})$ vanishes and we can
write the VEV of the commutator of the two currents as
\footnote{Here we assume (and later we show) that $G^{ab}_{1}(p^{2})=G^{ba}_{1}(p^{2})$.}
\begin{eqnarray}
\langle0|\left[J^{a}_{\mu}(x),J^{b}_{\nu}(y)\right]|0\rangle & = &
F^{ab}_{\mu\nu}(x-y) - F^{ab}_{\nu\mu}(y-x)  \nonumber \\
& = & \int{d^{2}p\: e^{-ip(x-y)}\left(p^{2}g_{\mu\nu}-
p_{\mu}p_{\nu}\right)G^{ab}_{1}(p^{2})\varepsilon(p^{0})}=  \nonumber
\end{eqnarray}
\begin{eqnarray}
&=&\int_{0}^{\infty}{dM^{2} \int{d^{2}p\: e^{-ip(x-y)}
\left(M^{2}g_{\mu\nu}-p_{\mu}p_{\nu}\right)G^{ab}_{1}(M^{2})
\varepsilon(p^{0})\delta(p^{2}-M^{2})}}  \nonumber \\
& = & 2\pi\int_{0}^{\infty}{dM^{2}\:G^{ab}_{1}(M^{2})(M^{2}g_{\mu\nu}+
\partial_{\mu}^{x}\partial_{\nu}^{x})\Delta_{2}(x-y,M^{2})},
\label{valid}
\end{eqnarray}
where
\begin{equation}
\varepsilon(p^{0})=\theta(p^{0})-\theta(-p^{0})=\frac{p^0}{|p^0|},
\end{equation}
and $\Delta_{2}(x-y,M^{2})$ is the Pauli-Jordan function in two
dimensions
\begin{equation}
\Delta_{2}(x-y,M^{2})= \frac{1}{2\pi} \int{d^{2}p\:e^{-ip(x-y)}
\delta(p^{2}-M^{2})\varepsilon(p^{0})}, \label{delta2}
\end{equation}
with the following familiar properties
\begin{eqnarray}
\partial_{0}^{x} \Delta_{2}(x,M^{2}) \left|_{x^{0}=0} \right.& = &
-i\delta(x^{1}),  \nonumber \\
\Delta_{2}(x,M^{2})\left|_{x^{0}=0}\right. & = & 0.\label{properties}
\end{eqnarray}
Using (\ref{delta2}) and (\ref{properties}) we can immediately write
\begin{equation}
\langle 0|\left[J^{a}_{0}(x),J^{b}_{0}(y)\right]|0\rangle_{E.T.}= \langle
0|\left[J^{a}_{1}(x),J^{b}_{1}(y)\right]|0\rangle_{E.T.}=0.
\end{equation}
For the combination $\mu=0$ and $\nu=1$ we find
\begin{equation}
\langle 0|\left[J^{a}_{0},J^{b}_{1}\right]|0\rangle_{E.T.}= 2\pi \int_{0}^{\infty}
{dM^{2}\:G^{ab}_{1}(M^{2})}\cdot\partial^{x}_{1} \delta(x^{1}-y^{1}).
\end{equation}
Interchanging the order of the integration in (\ref{valid}) is quite legal in our case
because both integrals exist. \footnote{For the 3+1 dimensional case this
property does not hold \cite{Kal}.}

What remains is to calculate the function $G^{ab}_{1}(p^{2})$. From (\ref{GG}%
) one finds
\begin{equation}
G^{ab}_{\mu\nu}(p)=\left(p^{2}g_{\mu\nu}-p_{\mu}p_{\nu}\right)
G^{ab}_{1}(p^{2}).
\end{equation}
Contracting both sides with $g^{\mu\nu}$, using (\ref{G}) and taking
into account that
only the state $|n\rangle$ containing one fermion-antifermion pair (as
the currents are {\it free}) contributes we get
\begin{eqnarray}
&{}&G^{ab}_{1}(p^{2}) = \frac{1}{p^{2}} \sum_{p_{1}+p_{2}}{\langle 0|
J^{a}_{\mu}|n\rangle \langle n| J^{b\ \mu} |0\rangle \nonumber} \\
{}\nonumber \\
&=& \frac{1}{(2\pi)^{2} p^{2}} \int{d^{2}p_{1}\:d^{2}p_{2}\:
\delta(p-p_{1}-p_{2})\delta(p_{1}^{2}-m^{2})
\delta(p_{2}^{2}-m^{2})\theta(p_{1}^{0})\theta(p_{2}^{0})} \times  \nonumber
\\
&{\ }&\ \ \ \ \ \ \ \ \ \ \ \times\;\mbox{Tr}\{\gamma_{\mu} (\not{p}%
_{1}+m)\gamma^{\mu}(\not{p}_{2}-m)\}\cdot\mbox{Tr}\{\tau^{a}\tau^{b}\}
\nonumber \\
\nonumber \\
&=& -\frac{1}{\pi^{2} p^{2}} \frac{m^2}{\sqrt{p^2(p^2-4m^{2})}}
\theta(p^{2}-4m^{2})\:\mbox{Tr}\{\tau^{a}\tau^{b}\},
\end{eqnarray}
where $m$ is the fermion mass. Because of
\footnote{This integral has evidently a singularity for $m=0$ which
compensates the factor of $m^2$; obviously, such an
"$m\cdot\frac{1}{m}$ efect" is completely analogous to that
arising in the dispersive derivation of the axial anomaly through
a relevant imaginary part (c.f. \cite{Adam}).}
\begin{equation}
m^{2}\int_{4m^{2}}^{\infty}{\frac{1}{a\sqrt{a(a-4m^{2})}}\:da}=\frac{1}{2}
\end{equation}
and the normalization
\begin{equation}
\mbox{Tr}\{\tau^{a}\tau^{b}\}=\frac{1}{2}\delta^{ab},
\end{equation}
we finally get
\begin{equation}
\langle 0|[J^{a}_{0}(x),J^{b}_{1}(y)]|0\rangle_{E.T.}= \frac{i}{2\pi}\delta^{ab}\,
\partial^{x}_{1}\delta(x^{1}-y^{1})
\end{equation}
and in the Abelian case
\begin{equation}
\langle 0|[J_{0}(x),J_{1}(y)]|0\rangle_{E.T.}= \frac{i}{\pi}%
\partial^{x}_{1}\delta(x^{1}-y^{1}).  \label{abel.ST}
\end{equation}


\section{Brandt's approach}


\label{3} \renewcommand{\theequation}{3.\arabic{equation}} %
\setcounter{equation}{0} We define the ETC of the two local
operators $A(x)$ and $B(y)$ as \cite{Bra}
\begin{equation}
[A(x),B(y)]_{E.T.}\equiv \lim_{\xi\to 0, \xi^{\prime}\to
0}[A(x,\xi),B(y,\xi^{\prime})] ,  \label{BA}
\end{equation}
where $A(x,\xi)$ and $B(y,\xi^{\prime})$ are functions of the renormalized
local operators $\psi (x)$, $\psi(x+\xi)$ evaluated at time $x_{0}$ and
\begin{equation}
A(x)=\lim_{\xi \to 0} A(x,\xi); \ \ \xi^{0}=0 .
\end{equation}
We define the current $J_{\Gamma}(x)$ as
\begin{equation}
J_{\Gamma}(x)\equiv\lim_{\xi\to 0} J_{\Gamma}(x,\xi),
\end{equation}
where
\begin{equation}
J_{\Gamma}(x,\xi)\equiv \frac{1}{2}\left[\bar{\psi}(x)\Gamma\psi(x+\xi)+
\bar{\psi}(x+\xi)\Gamma\psi(x)\right]
\end{equation}
is the point-split current and
\begin{equation}
\Gamma \in \{1,\gamma^{\mu},\cdots; \gamma^{\mu}\tau^{a},\cdots\}.
\end{equation}
Using the above definition and the relation
\begin{equation}
\{\psi_{\alpha a}(x),\psi^{+}_{\beta
b}(y)\}_{E.T.}=\delta_{\alpha\beta}\delta_{ab}\delta(x^{1}-{y}^{1})
\end{equation}
it is easy to find the useful identities
\begin{eqnarray}  \label{uzitecne rel.}
{[} \bar{\psi}(x)\Gamma\psi(y),\bar{\psi}(z) {]}_{E.T.} & = & \bar{\psi}%
(x)\Gamma\gamma^{0}\delta(y^{1}-z^{1}),  \nonumber \\
\nonumber \\
{[}\bar{\psi}(x)\Gamma\psi(y),\psi(z){]}_{E.T.}&=&
-\gamma^{0}\Gamma\psi(y)\delta(x^{1}-z^{1}),  \nonumber \\
\nonumber \\
{[} \bar{\psi}(x)\Gamma_{A}\psi(y), \bar{\psi}(z)\Gamma_{B}\psi(w) {]}%
_{E.T.} & = & \bar{\psi}(x)\Gamma_{A}\gamma^{0}\Gamma_{B}\psi(w)%
\delta(y^{1}-z^{1})-  \nonumber \\
&-&\bar{\psi}(z)\Gamma_{B}\gamma^{0}\Gamma_{A}\psi(y)\delta(x^{1}-w^{1}).
\nonumber \\
\end{eqnarray}
Now we can directly compute the single commutator of the two currents.
\begin{eqnarray}  \label{F}
&{}&{[}J_{\Gamma_{A}}(x,\xi),J_{\Gamma_{B}}(y,\xi^{\prime}){]}_{E.T.} =
\nonumber \\
&{}&  \nonumber \\
&=&\frac{1}{2} {\biggl[}\bar{\psi}(x)\Gamma_{A\stackrel{0}{}B}\psi(x+\xi)
\delta(x^{1}+\xi^{1}-y^{1})-\bar{\psi}(x)\Gamma_{B\stackrel{0}{}A}
\psi(x+\xi)\delta(x^{1}-y^{1})\ +  \nonumber \\
&\ &\ \ \ +\ \bar{\psi}(x+\xi)\Gamma_{A\stackrel{0}{}B}
\psi(x)\delta(x^{1}-y^{1})-\bar{\psi}(x+\xi)\Gamma_{B\stackrel{0}{}A}\psi(x)
\delta(x^{1}+\xi^{1}-y^{1}){\biggr]}  \nonumber \\
&=&\frac{1}{2}{\biggl[}\bar{\psi}(x)\Gamma_{A\stackrel{0}{}B}\psi(x+\xi)
\sum_{i=0}^{\infty}{(\xi^{1})^{i}\partial_{1}^{x\,(i)}}\delta(x^{1}-y^{1}) \ -
\nonumber \\
&\ &\ \ \ \ \ \ \ \ -\ \bar{\psi}(x)\Gamma_{B\stackrel{0}{}A}
\psi(x+\xi)\delta(x^{1}-y^{1})+  \nonumber \\
&\ &\ \ \ \ \ \ \ \ \ \ \ \ \ \ \ \ \ \ \ \ \ +\ \bar{\psi}(x+\xi)\Gamma_{A%
\stackrel{0}{}B} \psi(x)\delta(x^{1}-y^{1}) \ -  \nonumber \\
&\ &\ \ \ \ \ \ \ \ \ \ \ \ \ \ \ \ \ \ \ \ \ \ \ \ \ \ \ \ \ \ \ \ \ \ -\
\bar{\psi}(x+\xi)\Gamma_{B\stackrel{0}{}A}\psi(x) \sum_{i=0}^{\infty}({%
\xi^{1})^{i}\partial_{1}^{x\,(i)}}\delta(x^{1}-y^{1}) {\biggr]}  \nonumber \\
&=&\frac{1}{2}{\biggl[} \bar{\psi}(x)\Gamma_{[A\stackrel{0}{,}B]}\psi(x+\xi)
+\bar{\psi}(x+\xi)\Gamma_{[A\stackrel{0}{,}B]}\psi(x) {\biggr]}
\delta(x^{1}-y^{1})+  \nonumber \\
&\ &\ \ \ +\ \frac{1}{2} {\biggl[} \bar{\psi}(x)\Gamma_{A\stackrel{0}{}%
B}\psi(x+\xi)- \bar{\psi}(x+\xi)\Gamma_{B\stackrel{0}{}A}\psi(x){\biggr]}
\sum_{i=1}^{\infty}{(\xi^{1})^{i}\partial_{1}^{x\,(i)}}\delta(x^{1}-y^{1})
\nonumber \\
\nonumber \\
& = & J_{\Gamma_{[A\stackrel{0}{,}B]}}(x,\xi)\delta(x^{1}-y^{1})+  \nonumber
\\
&\ &\ \ \ +\ \frac{1}{2} {\bigg[}\bar{\psi}(x)\Gamma_{A\stackrel{0}{}%
B}\psi(x+\xi)- \bar{\psi}(x+\xi)\Gamma_{B\stackrel{0}{}A}\psi(x){\bigg]}
\sum_{i=1}^{\infty}{(\xi^{1})^{i}\partial_{1}^{x\,(i)}}\delta(x^{1}-y^{1}),
\nonumber \\
\end{eqnarray}
where the following notation was used
\begin{eqnarray}
\Gamma_{A\stackrel{0}{}B}&\equiv&\Gamma_{A}\gamma^{0}\Gamma_{B}, \\
\Gamma_{[A\stackrel{0}{,}B]}&\equiv&\Gamma_{A\stackrel{0}{}B}- \Gamma_{B%
\stackrel{0}{}A}
\end{eqnarray}
and the limit $\xi^{\prime}\to 0$ was already taken. It is easy to check
that the result does not depend on the order in which we take the limit
procedures.

The first term in (\ref{F}) (after taking the limit $\xi\to 0$) is, as we
could expect, the operator of the current.

Assuming about the Schwinger term that it is a c-number we find its form
by calculating the VEV of the second term in (\ref{F}).

Considering that
\begin{equation}
\lim_{\xi^{0} \to 0^{+}} \langle 0| \bar{\psi} (x)\Gamma_{A\stackrel{0}{}%
B}\psi(x+\xi)|0 \rangle = -\frac{i}{2\pi}\mbox{Tr}\Big\{\Gamma_{A\stackrel{0}{}B}%
\gamma_{1}\Big\} \frac{\xi^{1}}{(\xi^{1})^{2}-i\varepsilon},
\end{equation}
we find
\begin{eqnarray}  \label{s.com.1+1}
\lim_{\xi\to 0,\ \xi^{\prime}\to
0}[J_{\Gamma_{A}}(x,\xi),J_{\Gamma_{B}}(y,\xi^{\prime})]&=& J_{[\Gamma_{A}%
\stackrel{0}{,}\Gamma_{B}]}(y)\delta(x^{1}-y^{1})-  \nonumber \\
&-&\frac{i}{4\pi} \mbox{Tr}\Big\{\{\Gamma_{A}\stackrel{0}{,}%
\Gamma_{B}\}\gamma_{1}\Big\} \partial_{1}^{x}\delta(x^{1}-y^{1}).  \nonumber \\
\end{eqnarray}
Finally we get the following Schwinger terms (S.T.) for the different
combinations of the matrices
\begin{eqnarray}
\Gamma_{A}=\gamma_{0},\ \Gamma_{B}=\gamma_{1}\ \ \ \ \ \ \ \ \ \ \ &%
\mbox{S.T.}&=\frac{i}{\pi}\partial^{x}_{1}\delta(x^{1}-y^{1}),  \nonumber \\
\Gamma_{A}=\Gamma_{B}=\gamma_{0(1)}\ \ \ \ \ \ \ \ \ \ \ &\mbox{S.T.}&=0,
\nonumber \\
\Gamma_{A}=\tau^{a}\gamma_{0},\ \Gamma_{B}=\tau^{b}\gamma_{1}\ \ \ \ \ \ \ \
\ \ \ &\mbox{S.T.}&=\frac{i}{2\pi}\delta^{ab}
\partial^{x}_{1}\delta(x^{1}-y^{1}),  \nonumber \\
\Gamma_{A}=\tau^{a}\gamma_{0(1)},\ \Gamma^{B}=\tau^{b}\gamma_{0(1)}\ \ \ \ \
\ \ \ \ \ \ \ &\mbox{S.T.}&=0.
\end{eqnarray}
Comparing this with the previous section we observe that both
approaches - K\"allen's and Brandt's - give the same results.


\section{Anomaly and Schwinger terms}


\label{4} \renewcommand{\theequation}{4.\arabic{equation}}
\setcounter{equation}{0}

In this section I follow essencialy \cite{Gross}.

The massless Schwinger model has on the classical level the gauge respectively
the chiral symmetries from which the conservation of the vector respectively
the axial vector currents results
\begin{eqnarray}
\partial^{x}_{\mu}J^{\mu}(x)&=&0,  \nonumber \\
\partial^{y}_{\nu}J^{\nu}_{5}(y)&=&0,
\end{eqnarray}
where
\begin{equation}
J^{\nu}_{5}\equiv\bar{\psi}(x)\gamma^{\nu}\gamma^{5}\psi(x),
\end{equation}
where $\gamma^{5}=\gamma^{0}\gamma^{1}$.

Using in these equations and the canonical commutation relations (CCR)
we may formally derive identities like
\begin{eqnarray}
\partial^{x}_{\mu}\langle 0|T J^{\mu}(x)J^{\nu}_{5}(y) |0\rangle&=&0,
\nonumber \\
\partial^{y}_{\nu}\langle 0|T J^{\mu}(x)J^{\nu}_{5}(y) |0\rangle&=&0.
\end{eqnarray}
On the quantum level these equations - the vector respectively axial vector Ward
identities (VWI resp. AWI) - cannot be satisfied simultaneously
\cite{Gross} and then we speak about the anomaly respectively the anomalous Ward identities.

To show it we introduce a covariant two-point Green function
\begin{equation}
G^{\mu\nu}\equiv\langle 0|T^{*}J^{\mu}(x)J^{\nu}_{5}(y)|0\rangle.
\label{cov. Green}
\end{equation}
The $T^{*}$-product is defined for two Bose operators $A(x)$ and $B(y)$ as
\begin{equation}
T^{*} \big( A(x) B(y) \big) \equiv T \big( A(x) B(y);n) \big) + C(x,y;n) + c(x,y),
\end{equation}
where
\begin{equation}
T\big( A(x) B(y);n \big) \equiv \theta ([x-y]\cdot n)A(x)B(y)+ \theta
([y-x]\cdot n)B(y)A(x)
\end{equation}
is the generalized $T$-product, $n$ is a time-like vector, $n^{2}=1$,
$C(x,y;n)$ and $c(x,y)$ are the so-called contact or seagull terms.\footnote{
It is easy to see that for $n=(1,0)$ we get the ordinary $T$-product.}

To find $C(x,y;n)$ we use the property that $T^{*}$ is independent of $n$.
We get (for details see \cite{Gross})
\begin{equation}
C(x,y;n)=C(y;n)\delta^{2}(x-y) ,  \label{gen. seagull}
\end{equation}
where
\begin{equation}
C(y;n)=\int{dn^{\prime}_{\alpha}\:S^{\alpha}(y,n^{\prime})}
\end{equation}
and $S^{\alpha}(y,n^{\prime})$ is given by
\begin{equation}
\delta([x-y]\cdot n)\big[A(x),B(y)\big]=S(y;n)\delta^{2}(x-y)+S^{\alpha}
(y;n)P_{\alpha\beta}\partial^{\beta}\delta^{2}(x-y).
\end{equation}
$P_{\alpha\beta}$ is the spacelike projection operator
\begin{equation}
P_{\alpha\beta}=g_{\alpha\beta}-n_{\alpha}n_{\beta}.
\end{equation}
Note that $C(y;n)$ is determined up to an arbitrary function $C_{0}(y)$!

In the previous sections we have got the complete formulae for the
commutators and now we can rewrite them in a more compact and useful form
\begin{equation}
\delta([x-y]\cdot n)\big[J^{\mu}(x),J^{\nu}(y)\big]=S^{\mu\nu;\alpha}
(y;n)P_{\alpha\beta}\partial^{\beta}\delta^{2}(x-y) ,  \label{cov.com.}
\end{equation}
with
\begin{equation}
S^{\mu\nu;\alpha}(y;n)=S(y)(n^{\mu}g^{\nu\alpha}+n^{\nu}g^{\mu\alpha}).
\end{equation}
For 1+1 dimensions
\begin{equation}
J^{\mu}_{5}=\varepsilon^{\mu\nu}J_{\nu}
\end{equation}
and in case of (\ref{cov. Green}) it is easy to find that \footnote{%
The seagull $C(x,y;n)$ in (\ref{gen. seagull}) carries two indices $\mu, \nu$
now.}
\begin{equation}
C^{\mu\nu}(x,n)=S(x)n^{\mu}n^{\mu}\delta^{2}(x-y)+C_{0}^{\mu\nu}(x)
\end{equation}
and we get
\begin{equation}
\partial^{x}_{\mu}T^{*} \big( J^{\mu}(x)J^{\nu}_{5}(y) \big)= \big( C_{0\ \mu}^{\nu}(y)\partial^{\mu}_{x}+S(y)\partial^{\nu}_{x} \big) \delta^{2}(x-y).
\end{equation}
Now it is clear that if we want to preserve VWI we have to put
\begin{equation}
C_{0\ \mu\nu}(y)=-g_{\mu\nu}S(y)
\end{equation}
and then AWI is broken
\begin{equation}
\partial^{y}_{\nu}T^{*} \big( J^{\mu}(x)J^{\nu}_{5}(y) \big)=
S(y)\varepsilon^{\mu\nu}\partial^{y}_{\nu}\delta^{2}(x-y).
\end{equation}
On the contrary if we try to preserve AWI then VWI will be broken.

From the last equation the connection between the anomaly and the Schwinger
term is clear.

In our simple case the Schwinger term is reduced to (see (\ref{abel.ST}), (\ref{cov.com.}))
\[
S(x)=\frac{i}{\pi}
\]
and so AWI is given by
\begin{equation}
\partial^{y}_{\nu}T^{*} \big( J^{\mu}(x),J^{\nu}_{5}(y) \big)=\frac{i}{\pi}
\varepsilon^{\mu\nu}\partial^{y}_{\nu}\delta^{2}(x-y).
\end{equation}


\section{Double commutators in K\"allen's approach}

\label{5} \renewcommand{\theequation}{5.\arabic{equation}} %
\setcounter{equation}{0}
In the calculation of the double commutators we proceed similarly as
in Section 2.\\
Let us consider the following double commutator
\begin{equation}
\Big[[J_{A}(x),J_{B}(y)],J_{C}(z)\Big],
\end{equation}
later with the specific values of $A,\ B$ and $C$:
\begin{equation}
A=\gamma_{1}\tau^{a},\ \ B=\gamma_{1}\tau^{b},\ \ C=\gamma_{1}\tau^{c}.
\end{equation}
We introduce the function
\begin{eqnarray}
F_{\alpha\beta\gamma}^{abc}&=&
\sum_{n,\ m}\langle 0|J_{\alpha}^{a} (x)|n\rangle
            \langle n|J_{\beta}^{b} (y)|m\rangle
            \langle m|J_{\gamma}^{c} (z)|0\rangle \nonumber \\
&=&
\sum_{n,\ m}\langle 0|J_{\alpha}^{a}|n\rangle
            \langle n|J_{\beta}^{b}|m\rangle
            \langle m|J_{\gamma}^{c}|0\rangle
            e^{-ip^{(n)}(x-y)}e^{-iq^{(m)}(y-z)} \nonumber \\
&=&\int {
d^{2}p\,d^{2}q\:
e^{-ip(x-y)}e^{-iq(y-z)}\theta(p^{0})\theta(q^{0})
G_{\alpha \beta \gamma}^{abc}(p,q)},
\end{eqnarray}
where
\begin{eqnarray}
G_{\alpha \beta \gamma}^{abc}(p,q)&=&
\sum_{n,\ m}\delta^{2}(p^{(n)}-p)\delta^{2}(q^{(m)}-q)
\langle 0|J_{\alpha}^{a}|n\rangle
\langle n|J_{\beta}^{b}|m\rangle
\langle m|J_{\gamma}^{c}|0\rangle \\
&\stackrel{eff}{=}&
\frac{i\pi f^{abc}}{8}
\int\frac{d^{2}r\,d^{2}s\,d^{2}u}{\pi^{4}}
\theta(r^{0})\theta(s^{0})\theta(u^{0})
\times \nonumber \\
&\times&
\delta(r^{2}-m^{2})\delta(s^{2}-m^{2})
\delta(u^{2}-m^{2})\delta^{2}(p-r-s)\delta^{2}(q-s-u)\times\nonumber\\
&\times&(r_{1}s_{0}u_{0}+r_{0}s_{1}u_{0}+r_{0}s_{0}u_{1}).
\end{eqnarray}
The abbreviation $eff$ means the effective equality, i.e. we drop terms which do not contribute to the equal time limit.
With this remark in mind
\begin{eqnarray}
F_{\alpha\beta\gamma}^{abc}&=&\frac{i\pi f^{abc}}{8}
\int
d^{2}p\,d^{2}q
\:e^{-ip(x-y)}e^{-iq(y-z)}
\theta(p^{0})\theta(q^{0})\times\nonumber\\
&{}&\nonumber\\
&\times&
\int
\frac{d^{2}r\,d^{2}s\,d^{2}u}{\pi^{4}}
\theta(r^{0})\theta(s^{0})\theta(u^{0})
\delta(r^{2}-m^{2})\delta(s^{2}-m^{2})
\delta(u^{2}-m^{2})\times\nonumber\\
&{}&\nonumber\\
&{}&\times\ \delta^{2}(p-r-s)\delta^{2}(q-s-u)
(r_{1}s_{0}u_{0}+r_{0}s_{1}u_{0}+r_{0}s_{0}u_{1}).
\end{eqnarray}
Using notation
\begin{equation}
[ts]=\theta(r^{0})\theta(s^{0})\theta(u^{0})\delta(r^{2}-m^{2})\delta(s^{2}-m^{2})
\delta(u^{2}-m^{2})(r_{1}s_{0}u_{0}+r_{0}s_{1}u_{0}+r_{0}s_{0}u_{1}),
\end{equation}
we can write
\begin{eqnarray}
\Big[[J_{1}^{a}(x),J_{1}^{b}(y)],J_{1}^{c}(z)\Big]=
\frac{i\pi
f^{abc}}{8}\int\frac{d^{2}r\,d^{2}s\,d^{2}u}{\pi^{4}}\ [ts]
\times\ \ \ \ \ \ \ \ \ \ \ \ \ \ \ \ \ \ \ \ \ \ \ \ \ \ &{}&\nonumber\\
\times\ \Big(e^{-ir(x-y)}e^{-is(x-z)}e^{-iu(y-z)}
+e^{-ir(y-x)}e^{-is(y-z)}e^{-iu(x-z)}\ -&{}&\nonumber\\
-\ e^{-ir(z-x)}e^{-is(z-y)}e^{-iu(x-y)}-e^{-ir(z-y)}e^{-is(z-x)}e^{-iu(y-x)}
\Big)
&=&\nonumber
\end{eqnarray}
\begin{eqnarray}
&=&
\frac{i\pi f^{abc}}{8}
\int
\frac{d^{2}r\,d^{2}s\,d^{2}u}{\pi^{4}}\:[ts]\ \times \nonumber \\
&\times&\Bigg[e^{-ir(x-y)}e^{-is(x-z)}e^{-iu(y-z)}
\Big(\theta(r^{0})\theta(s^{0})\theta(u^{0})
\theta(r^{0}+s^{0})\theta(s^{0}+u^{0})\ -\nonumber\\
&{}& \ \ \ \ \ \ \ \ \ \ \ \ \ \ \ \ \ \ \ \ \ \ \ \
-\ \theta(-r^{0})\theta(s^{0})\theta(u^{0})
\theta(-r^{0}+s^{0})\theta(s^{0}+u^{0})\Big)\ -\nonumber\\
&{}& \nonumber \\
&-&\ \  e^{-ir(z-x)}e^{-is(z-y)}e^{-iu(x-y)}
\Big(\theta(r^{0})\theta(s^{0})\theta(u^{0})
\theta(r^{0}+s^{0})\theta(s^{0}+u^{0})\ -\nonumber\\
&{}& \ \ \ \ \ \ \ \ \ \ \ \ \ \ \ \ \ \ \ \ \ \ \ \
-\ \theta(r^{0})\theta(s^{0})\theta(-u^{0})
\theta(r^{0}+s^{0})\theta(s^{0}-u^{0})\Big)
\Bigg]\\
&=&\frac{i\pi f^{abc}}{8}
\int
\frac{d^{2}r\,d^{2}s\,d^{2}u}{\pi^{4}}\ [ts]\times\nonumber\\
&{}&\ \ \ \ \ \ \ \ \ \ \ \times e^{-ir(x-y)}e^{-is(x-z)}e^{-iu(y-z)}
\varepsilon(r^{0})\theta(s^{0})\theta(u^{0})\theta(u^{0}+s^{0})
-\nonumber\\
&{}&\nonumber\\
&-&\frac{i\pi f^{abc}}{8}
\int
\frac{d^{2}r\,d^{2}s\,d^{2}u}{\pi^{4}}\ [ts]\times\nonumber\\
&{}&\ \ \ \ \ \ \ \ \ \ \ \times e^{-ir(z-x)}e^{-is(z-y)}e^{-iu(x-y)}
\varepsilon(u^{0})\theta(r^{0})\theta(s^{0})\theta(r^{0}+s^{0})\nonumber\\
\\
&=&\frac{i\pi f^{abc}}{4}
\int
\frac{d^{2}r}{2\pi}
e^{-ir(x-y)}\varepsilon(r^{0})
r_{0}\delta(r^{2}-m^{2})
\int\frac{d^{2}s\,d^{2}u}{\pi^{3}}(s_{0}u_{1}+s_{1}u_{0})\times\nonumber\\
&{}&\times\ \theta(s^{0})\theta(u^{0})\theta(u^{0}+s^{0})
\delta(s^{2}-m^{2})\delta(u^{2}-m^{2})
e^{-is(x-z)}e^{-iu(y-z)}-\nonumber\\
&{}&\nonumber\\
&-&\frac{i\pi f^{abc}}{4}
\int
\frac{d^{2}u}{2\pi}
e^{-iu(x-y)}\varepsilon(u^{0})
u_{0}\delta(u^{2}-m^{2})
\int\frac{d^{2}r\,d^{2}s}{\pi^{3}}(r_{0}s_{1}+r_{1}s_{0})\times\nonumber\\
&{}&\times\ \theta(r^{0})\theta(s^{0})\theta(r^{0}+s^{0})
\delta(r^{2}-m^{2})\delta(s^{2}-m^{2})
e^{-ir(z-x)}e^{-is(z-y)}\\
\nonumber\\
&\stackrel{E.T.}{=}&\frac{if^{abc}}{2}\delta(x-y)
\int\frac{d^{2}t}{2}(e^{-it(z-x)}-e^{-it(x-z)})\theta(t^{0})\times\nonumber\\
&\times&\int\frac{d^{2}r\,d^{2}s}{\pi^{3}}(r_{0}s_{1}+r_{1}s_{0})
\theta(r^{0})\theta(s^{0})\delta(r^{2}-m^{2})\delta(s^{2}-m^{2})
\delta^{2}(t-r-s)\nonumber\\
&{}&\\
&=&\frac{if^{abc}}{2\pi}\delta(x^{1}-y^{1})\partial_{1}^{y}\delta(y-z),
\end{eqnarray}
where in the last equality we used (\ref{useful}) from the Appendix.
The procedures of the calculation of the remaining double commutators are practically the same and we can omit them.

In addition, in the following section we derive the general form (\ref{gen.ST}) of
the VEV of the double commutator of the free current using a simpler formalism.


\section{Double commutators in Brandt's approach}

\label{6} \renewcommand{\theequation}{6.\arabic{equation}}
\setcounter{equation}{0}
Here we consider two ways how to calculate the
double commutators using Brandt's approach. One of them (I will call it the "w-way") could violate the Jacobi identity as was shown by Banerjee, Rothe \& Rothe \cite{Ban}.

We start from the general form of the single commutator, which we got in the
previous sections
\begin{equation}
[J_{A}(x),J_{B}(y)]_{E.T.}=\delta(x^{1}-y^{1})J_{[A\stackrel{0}{,}B]}(y)+
S_{AB}\partial_{1}^{x}\delta(x^{1}-y^{1}).  \label{gen. singl}
\end{equation}
We define the double commutator similarly as the single one (see (\ref{BA}))
\begin{equation}
\big[ [J_{A}(x),J_{B}(y)],J_{C}(z) \big]_{E.T.} \equiv \lim_{\xi\to 0,
\xi^{\prime}\to 0, \xi^{\prime\prime}\to 0} \big[ [J_{A}(x,\xi),J_{B}(y,%
\xi^{\prime})],J_{C}(z,\xi^{\prime\prime}) \big]_{E.T.},  \label{gen. double}
\end{equation}
where we use $A$, resp. $B$, resp. $C$ as the shorthand notation for
$\Gamma_{A}$, resp. $\Gamma_{B}$, resp. $\Gamma_{C}$.


\subsection{W-way}


\label{6.1} \renewcommand{\theequation}{6.1.\arabic{equation}} %
\setcounter{equation}{0}

The "w-way" consists in the assumption that we can put the limit
procedures inside the external commutator i.e.
\begin{equation}
\lim_{\xi^{\prime\prime}\to 0} \big[ \lim_{\xi \to 0, \xi^{\prime}\to 0}
[J_{A}(x,\xi),J_{B}(y,\xi^{\prime}) ]_{E.T.},J_{C}(z,\xi^{\prime\prime}) %
\big]_{E.T.}
\end{equation}
and then use (\ref{gen. singl}). Employing this procedure we can write
\begin{equation}
\big[ [J_{A}(x),J_{B}(y)],J_{C}(z) \big]_{E.T.}=\big[ \delta(x^{1}-y^{1})J_{[A%
\stackrel{0}{,}B]}(y)+ S_{AB}\partial_{1}^{x}\delta(x^{1}-y^{1}),J_{C}(z) \big]%
_{E.T.}.
\end{equation}
Assuming that $S_{AB}$ is a c-number we find
\begin{eqnarray}  \label{wrong way result}
\big[ [J_{A}(x),J_{B}(y)],J_{C}(z) \big]_{E.T.}&=&\delta(x^{1}-y^{1})
\delta(y^{1}-z^{1}) J_{\big[
[A\stackrel{0}{,}B]\stackrel{0}{,}C\big]}(z)-
\nonumber \\
&\ &-\ \frac{i}{4\pi}S_{AB,C}\delta(x^{1}-y^{1})\partial_{1}^{y}
\delta(y^{1}-z^{1}),  \nonumber \\
\end{eqnarray}
where we introduced the notation
\begin{equation}
S_{AB,C}=\mbox{Tr}\bigg\{\big\{[A\stackrel{0}{,}B] \stackrel{0%
}{,}C\big\}\gamma_{1}\bigg\}.\label{ST for double}
\end{equation}
For
\begin{equation}
A=\tau^{a}\gamma^{\alpha},\ \ B=\tau^{b}\gamma^{\beta},\ \
C=\tau^{c}\gamma^{\gamma} \label{priklad}
\end{equation}
we obtain
\[
\langle 0|\big[ [J_{A}(x),J_{B}(y)],J_{C}(z) \big]_{E.T.}|0\rangle=-\frac{1}{2\pi}f^{abc}\times \ \ \ \ \ \ \ \ \ \ \ \ \ \ \ \ \ \ \ \
\ \ \ \ \ \ \ \ \ \ \ \ \ \ \ \ \ \ \ \ \ \ \ \ \ \ \ \ \ \ \ \ \ \ \ \ \ \
\]
\[
\times\big(g^{\alpha 1}g^{\beta 1}g^{\gamma 1} +g^{\alpha 0}g^{\beta
0}g^{\gamma 1} +g^{\alpha 0}g^{\beta 1}g^{\gamma 0} +g^{\alpha 1}g^{\beta
0}g^{\gamma 0}\big)
\delta(x^{1}-y^{1})\partial_{1}^{y} \delta(y^{1}-z^{1}),
\]
\begin{equation}  \label{gen.ST}
\end{equation}
where
\begin{equation}
f^{abc}=-2i\mbox{Tr}\Big\{[\tau^a,\tau^{b}]\tau^{c}\Big\}.
\end{equation}


\subsection{C-way}


\label{6.2} \renewcommand{\theequation}{6.2.\arabic{equation}} %
\setcounter{equation}{0}

Using (\ref{uzitecne rel.}) one finds
\[
\big[
[\bar{\psi}(x)A\psi(y),\bar{\psi}(z)B\psi(w)],\bar{\psi}(u)C\psi(v) \big]%
_{E.T.}=\ \ \ \ \ \ \ \ \ \ \ \ \ \ \ \ \ \ \ \ \ \ \ \ \ \ \ \ \ \
\]
\begin{eqnarray}
&=&\ \ \ \bar{\psi}(x)\overline{ABC}\psi(v)\delta(w^{1}-u^{1})%
\delta(y^{1}-z^{1})-  \nonumber \\
&{}&-\ \bar{\psi}(u)\overline{CAB}\psi(w)\delta(x^{1}-v^{1})%
\delta(y^{1}-z^{1})-  \nonumber \\
&{}&-\ \bar{\psi}(z)\overline{BAC}\psi(v)\delta(y^{1}-u^{1})%
\delta(x^{1}-w^{1})+  \nonumber \\
&{}&+\ \bar{\psi}(u)\overline{CBA}\psi(y)\delta(z^{1}-v^{1})
\delta(x^{1}-w^{1}),  \label{correct way}
\end{eqnarray}
where for the case (\ref{priklad})
\begin{equation}
\overline{ABC}=\tau^{a}\tau^{b}\tau^{c}
\gamma^{\alpha}\gamma^{0}\gamma^{\beta}\gamma^{0}\gamma^{\gamma},\ \
\overline{CAB}=\tau^{c}\tau^{a}\tau^{b}
\gamma^{\gamma}\gamma^{0}\gamma^{\alpha}\gamma^{0}\gamma^{\beta},
\dots\ .
\end{equation}
What is really important is to take one of the two internal limits as the
last one, i.e. $\xi$ or $\xi^{\prime}$, otherwise we immediately go back to
the "w-way". Concretely in (\ref{correct way}) we can put
\begin{equation}
x=x,\ y=x+\xi,\ z=w=y,\ u=v=z.
\end{equation}

It is now straightforward but tedious to show that the result is the same as
that obtained by the "w-way" (see (\ref{wrong way result}), (\ref{gen.ST})).

Actually, in the case of 1+1 dimensions it is a little bit redundant to speak
about two ways because both give the same result. But in the following
section we will see their possible difference.

Concluding this section we can say that both methods, i.e. Brandt's and
K\"allen's, give the same results for the form of the Schwinger terms in the
double commutators of the free currents.


\section{Failure of Jacobi identity?}


\label{7} \renewcommand{\theequation}{7.\arabic{equation}} %
\setcounter{equation}{0} In the previous section we have got the following
general formula for the double commutator
\begin{eqnarray}
\big[ [J_{A}(x),J_{B}(y)],J_{C}(z) \big]_{E.T.}&=&
\delta(x^{1}-y^{1})\delta(y^{1}-z^{1}) J_{\big[ [A\stackrel{0}{,}B]\stackrel{%
0}{,}C\big]}(z)-  \nonumber \\
&\ &-\ \frac{i}{4\pi}S_{AB,C}\delta(x^{1}-y^{1})\partial_{1}^{y} \delta(y^{1}-z^{1}),
\nonumber \\
\end{eqnarray}
where the form of $S_{AB,C}$ is given by (\ref{ST for double}).

Now using the identities
\begin{eqnarray}
\delta(x^{1}-y^{1})\delta(y-z) J_{\big[ [A\stackrel{0}{,}B]\stackrel{0}{,}C%
\big]}(z) &+&  \nonumber \\
+\delta(y^{1}-z^{1})\delta(z-x) J_{\big[ [B\stackrel{0}{,}C]\stackrel{0}{,}A%
\big]}(x) &+&  \nonumber \\
+\delta(z^{1}-x^{1})\delta(x-y) J_{\big[ [C\stackrel{0}{,}A]\stackrel{0}{,}B%
\big]}(y) &=& 0\ ,
\end{eqnarray}
\begin{eqnarray}
\delta(x^{1}-y^{1})\partial_{1}^{y}\delta(y^{1}-z^{1}) &+&  \nonumber \\
+\delta(y^{1}-z^{1})\partial_{1}^{z}\delta(z^{1}-x^{1}) &+&  \nonumber \\
+\delta(z^{1}-x^{1})\partial_{1}^{x}\delta(x^{1}-y^{1}) &=& 0
\label{triv. delta}
\end{eqnarray}
and
\begin{equation}
S_{AB,C}=S_{BC,A}=S_{CA,B}\ ,
\end{equation}
it is easy to see that the Jacobi identity is fulfilled.

The Jacobi identity for the free currents cannot be broken in 1+1 dimensions
because both ways - "w" and "c" - leads to the same results.

As an illustration that it does not have to be always true, I mention an example given
in \cite{Ban} in which these
two procedures {\it do give} different results.

We consider the Jacobi identity for the operators $J^{i}({x}),\ J^{j}({y}),\
J^{0}_{5}({z})$ in 3+1 dimensions.\footnote{%
It means that the formula (\ref{s.com.1+1}) is not valid any more, but we can simply
generalize the method for the 3+1 dimensional case \cite{Bra}.} \\
Because of
\begin{eqnarray}
[J^{0}_{5}({x}),J^{j}({y})]_{E.T.}&=&0,\\
\nonumber \\
{[}J^{i}(y),J^{j}(z)]_{E.T.}&=&2i
\varepsilon^{ijk}J_{5\ k}(y)\delta^{3}({\bf y}-{\bf z}),\\
\nonumber \\
{[}J^{0}_{5}({x}),J^{i}({y})]_{E.T}&=&\frac{2i}{\pi^{2}}
\bigg[\frac{1}
{
\mbox{\boldmath{$\xi$}}
^{2}}C_{2}+\frac{1}{8}C_{4}{\bf
\nabla^{2}}\biggr]\partial^{i}\delta^{3}({\bf x}-{\bf y}),
\end{eqnarray}
then using the "w-way" we find
\begin{eqnarray}
\big[J^{0}_{5}({x}), [J^{i}({y}),J^{j}({z})] \big]_{E.T.}&=&-\frac{4}{\pi^{2}%
} \varepsilon^{ijk}\bigg[
\frac{1}
{
\mbox{\boldmath{$\xi$}}
^{2}}C_{2}
+\frac{1}{8}C_{4}{\bf \nabla^{2}}\biggr] %
\partial_{k} \delta^{3}({\bf x}-{\bf y})\delta^{3}({\bf y}-{\bf z}),
\nonumber \\
\\
\big[V^{i}({y}), [J^{j}({z}),J^{0}_{5}({x})] \big]_{E.T.}&=&0, \label{prvni blbej}\\
\big[V^{j}({z}), [J^{0}_{5}({x}),J^{i}({y})] \big]_{E.T.}&=&0  \label{druhy blbej}
\end{eqnarray}
and the Jacobi identity is evidently broken.
The constants $C_{2}$ and $C_{4}$ are given by the equations
\begin{eqnarray}
\langle \frac{\xi^{i} \xi^{j}}
{
\mbox{\boldmath{$\xi$}}
^{2}}
\rangle &=& C_{2}\delta^{ij},\\
\langle \frac{\xi^{i} \xi^{j} \xi^{k} \xi^{l}}
{
(
\mbox{\boldmath{$\xi$}}
^{2})^{2}
}\rangle &=& C_{4}(
\delta^{ij}\delta^{kl}+\delta^{ik}\delta^{jl}+\delta^{il}\delta^{jk}),
\end{eqnarray}
where $\langle\ \rangle$ means the average over spatial directions.

When one takes the internal limit procedures at the end of the calculation,
the results for the double commutators (\ref{prvni blbej}) and (\ref{druhy blbej})
are different \cite{Ban}
\begin{eqnarray}
&{}&\big[J^{i}({y}), [J^{j}({z}),J^{0}_{5}({x})] \big]_{E.T.}= -\frac{2%
}{\pi^{2}}\varepsilon^{ijk} \bigg[ \frac{2C_{2}}{{\bf \varepsilon}^{2}}
\delta^{3}({\bf x}-{\bf y}) \partial_{k}\delta^{3}({\bf y}-{\bf z})+
\nonumber \\
&{+}&C_{4}\bigg( \frac{1}{4}\partial_{k}\delta^{3}({\bf y}-{\bf z}%
)\nabla^{2} \delta^{3}({\bf x}-{\bf y})+\frac{1}{2}\partial_{l}\delta^{3}(%
{\bf y} -{\bf z})\cdot\partial_{l} \partial_{k}\delta^{3}({\bf x}-{\bf y})-
\nonumber \\
&{-}&\partial_{l}\partial_{k}\delta^{3}({\bf y}-{\bf z})\cdot\partial_{l}
\delta^{3}({\bf x}-{\bf y}) - \frac{1}{2}\nabla^{2}\delta^{3}({\bf y}-{\bf z}%
)\cdot\partial_{k}\delta^{3} ({\bf x}-{\bf y})+  \nonumber \\
&+&\nabla^{2}\partial_{k}\delta^{3}({\bf y}-{\bf z}) \cdot\delta^{3}({\bf x}-%
{\bf y})\bigg)\bigg], \\
\nonumber \\
\nonumber \\
&{}&\big[J^{0}_{5}({x}), [J^{i}({y}),J^{j}({z})] \big]_{E.T.}= \frac{2}{%
\pi^{2}}\varepsilon^{ijk} \bigg[ \frac{2C_{2}}{{\bf \varepsilon}^{2}}
\partial_{k}\delta^{3}({\bf x}-{\bf y})\delta^{3}({\bf y}-{\bf z})+
\nonumber \\
&{+}&\frac{C_{4}}{6}\bigg(3\nabla^{2}\partial_{k}\delta^{3}({\bf x}-{\bf y})
\delta^{3}({\bf y}-{\bf z})+ \partial_{k}\delta^{3}({\bf x}- {\bf y}%
)\nabla^{2}\delta^{3}({\bf y}-{\bf z})+  \nonumber \\
&+& 2\partial_{l}\partial_{k}\delta^{3}({\bf y}-{\bf z})\cdot\partial_{l}
\delta^{3}({\bf x}-{\bf y}) -\nabla^{2}({\bf x}-{\bf y})\partial_{k}%
\delta^{3}({\bf y}-{\bf z})  \nonumber \\
&-& 2\partial_{l}\partial_{k}\delta^{3}({\bf x}-{\bf y})\cdot
\partial_{l}\delta^{3}({\bf y}-{\bf z})\bigg)\bigg]
\end{eqnarray}
and the Jacobi identity is fulfilled.


\section{Conclusion}


\label{8} \renewcommand{\theequation}{8.\arabic{equation}} %
\setcounter{equation}{0}
I considered the single and double commutators of free currents in a
1+1 dimensional quantum field theory. In order to shed more light on their properties I used two methods to perform the calculation, K\"allen's and Brandt's, which are mathematically rigorous and elegant.

Using these results I determined the anomalous axial vector
Ward identity and showed the correct way of the regularization of the
double commutators which does not violate the Jacobi identity.

I hope that the statement from Introduction was clarified in this paper
and I just work on improvements of point-splitting (Brandt's) method
in higher dimensions.


\section*{Acknowledgements}


I would like to thank Prof. Bertlmann for giving me this interesting
problem and to Prof. J. Ho\v rej\v s\'{\i} for useful discussions.
My thanks belong to both of the two as well as Dr. Ch. Adam for
their help with completion of this article.
Also I am grateful to Dr. J.Novotn\'y, Mgr. M. Schnabl for helpful
discussions and to Dr. M. St\"ohr for computer help.


\section*{Appendix}


\label{A} \renewcommand{\theequation}{A.\arabic{equation}} %
\setcounter{equation}{0}

Here we derive the useful identity which is needed in the calculation of
the double commutator.\footnote{For the notation see Section 2.}\\
Firstly we consider the single commutator of the two currents
$J_{\mu}^{a}$ and $J_{\nu}^{b}$
\begin{equation}
\langle0|\left[J^{a}_{\mu}(x),J^{b}_{\nu}(y)\right]|0\rangle
=\int{d^{2}p\: e^{-ip(x-y)}\theta(p^{0})G^{ab}_{\mu\nu}(p^{2})},
\end{equation}
where
\begin{eqnarray}
G^{ab}_{\mu\nu}(p^{2})&=&\frac{1}{4}\int\frac{d^{2}r\,d^{2}s}{\pi^{2}}
\theta(r^{0})\theta(s^{0})\delta(r^{2}-m^{2})\delta(s^{2}-m^{2})
\delta^{2}(t-r-s)\times\nonumber\\
&{}&\ \ \ \times\ \mbox{Tr}\{\gamma_{\mu} (\not{r}+m)\gamma_{\nu}(\not{s}-m)\}
\mbox{Tr}\{\tau^{a}\tau^{b}\}.
\end{eqnarray}
For $\mu=0$ and $\nu=1$ we get
\begin{eqnarray}
G^{ab}_{\mu\nu}(p^{2})&=&\frac{1}{2}\int\frac{d^{2}r\,d^{2}s}{\pi^{2}}
\theta(r^{0})\theta(s^{0})\delta(r^{2}-m^{2})\delta(s^{2}-m^{2})
\delta^{2}(t-r-s)\times\nonumber\\
&{}&\ \ \ \times\ (r_{0}s_{1}+r_{1}s_{0})\mbox{Tr}\{\tau^{a}\tau^{b}\}.
\end{eqnarray}
Finally
\begin{eqnarray}
\langle0|\left[J^{a}_{\mu}(x),J^{b}_{\nu}(y)\right]|0\rangle &=&
\frac{1}{2}\int\frac{d^{2}t}{2}(e^{-it(z-x)}-e^{-it(x-z)})
\theta(t^{0})\times\nonumber\\
&{}&\nonumber\\
&{}&\times\ \int\frac{d^{2}r\,d^{2}s}{\pi^{3}}(r_{0}s_{1}+r_{1}s_{0})
\theta(r^{0})\theta(s^{0})\times\nonumber\\
&{}&\nonumber\\
&{}&\times\ \delta(r^{2}-m^{2})\delta(s^{2}-m^{2})
\delta^{2}(t-r-s)\mbox{Tr}\{\tau^{a}\tau^{b}\}.\nonumber\\
\end{eqnarray}
For the other side we have from Section 2
\begin{equation}
\langle 0|[J_{0}^{a}(x),J_{1}^{b}(y)]|0\rangle_{E.T.}=
\frac{i}{\pi}\partial^{x}_{1}\delta(x^{1}-y^{1})\mbox{Tr}\{\tau^{a}\tau^{b}\}.
\end{equation}
So we obtain the relation
\begin{eqnarray}
\frac{1}{2}\int\frac{d^{2}t}{2}(e^{-it(z-x)}-e^{-it(x-z)})\theta(t^{0})
\int\frac{d^{2}r\,d^{2}s}{\pi^{3}}(r_{0}s_{1}+r_{1}s_{0})&\times&\nonumber\\
\times\ \theta(r^{0})\theta(s^{0})\delta(r^{2}-m^{2})\delta(s^{2}-m^{2})
\delta^{2}(t-r-s)\nonumber&=&\\
&=&\!\!\!\!\!\frac{i}{\pi}\partial^{x}_{1}\delta(x^{1}-y^{1}).\label{useful}
\nonumber\\
\end{eqnarray}


\end{document}